\documentclass{article}

\usepackage{microtype}
\usepackage{graphicx}
\usepackage{subfigure}
\usepackage{booktabs} 
\usepackage[english]{babel}
\usepackage{hyperref}

\usepackage[accepted]{icml2023}

\usepackage{amsmath}
\usepackage{amssymb}
\usepackage{mathtools}
\usepackage{amsthm}
\usepackage[capitalize,noabbrev]{cleveref}

\theoremstyle{plain}

\theoremstyle{definition}

\theoremstyle{remark}

\usepackage[]{todonotes}
\usepackage{graphicx}
\usepackage{xcolor}
\usepackage{xspace}
\usepackage{enumitem}

\newcommand{\ours}{SoundStorm\xspace}
\newcommand{\soundstream}{SoundStream\xspace}
\newcommand{\audiolm}{AudioLM\xspace}
\newcommand{\spear}{SPEAR-TTS\xspace}

\icmltitlerunning{\ours: Efficient Parallel Audio Generation}

\begin{document}

\twocolumn[
\icmltitle{\ours: Efficient Parallel Audio Generation}



\icmlsetsymbol{equal}{*}

\begin{icmlauthorlist}
\icmlauthor{Zal\'an Borsos}{google}
\icmlauthor{Matt Sharifi}{google}
\icmlauthor{Damien Vincent}{google}
\icmlauthor{Eugene Kharitonov}{google}
\icmlauthor{Neil Zeghidour}{google}
\icmlauthor{Marco Tagliasacchi}{google}
\end{icmlauthorlist}

\icmlaffiliation{google}{Google Research}

\icmlcorrespondingauthor{Zal\'an Borsos}{zborsos@google.com}

\icmlkeywords{Machine Learning, ICML}

\vskip 0.3in
]



\printAffiliationsAndNotice{}  

\begin{abstract}
We present \ours, a model for efficient, non-autoregressive audio generation. \ours{} receives as input the semantic tokens of \audiolm, and relies on bidirectional attention and confidence-based parallel decoding to generate the tokens of a neural audio codec. Compared to the autoregressive generation approach of \audiolm, our model produces audio of the same quality and with higher consistency in voice and acoustic conditions, while being two orders of magnitude faster. \ours generates 30 seconds of audio in 0.5 seconds on a TPU-v4. We demonstrate the ability of our model to scale audio generation to longer sequences by synthesizing high-quality, natural dialogue segments, given a transcript annotated with speaker turns and a short prompt with the speakers' voices.  Audio samples are available at \url{https://google-research.github.io/seanet/soundstorm/examples/}
\end{abstract}

\section{Introduction}

\looseness=-1
Modeling discrete representations of audio produced by neural codecs~\citep{soundstream, defossez2022high} makes the task of audio generation amenable to the powerful Transformer-based sequence-to-sequence modeling approaches~\cite{attentionvaswani}. 
Casting unconditional and conditional audio generation as sequence-to-sequence modeling has unlocked rapid progress in speech continuation~\citep{audiolm}, text-to-speech~\citep{valle,speartts}, and general audio and music generation~\citep{audiogen, musiclm}.

\looseness=-1
For generating high-quality audio by modeling the tokens of a neural codec, the rate of the discrete representation must be increased, resulting in either an exponential growth in codebook size or in long token sequences. While the exponential growth of the codebook is prohibitive due to memory limitations, in turn, long token sequences also present computational challenges for autoregressive models. In particular, attention-based models, which are the main focus of this work, will incur quadratic runtime complexity with respect to the sequence length for calculating the self-attention. Thus, addressing the trade-off between perceptual quality and runtime is one of the core challenges for audio generation. 

\looseness=-1
The problem of generating long audio token sequences can be addressed by at least three orthogonal approaches, or a combination thereof: i) efficient attention mechanisms \citep{reformer,performer,nystromformer,perceiverar}, ii) non-autoregressive, parallel decoding schemes \citep{nar-nmt,maskpredict,chang2022maskgit}, iii) custom architectures adapted to the special structure of the tokens produced by neural audio codecs~\citep{audiogen, valle, lee2022}. However, in the context of modeling the token sequence of neural audio codecs, either unconditionally or based on weak conditioning such as text, the efficient generation of  long, high-quality audio segments remains an open problem. 

We believe that it is the special structure of the audio token sequence that holds the most promise for future advances in long-sequence audio modeling. Concretely, both \soundstream~\citep{soundstream} and EnCodec~\citep{defossez2022high} rely on Residual Vector Quantization (RVQ), where each compressed audio frame is quantized by a series of quantizers, with each quantizer operating on the residual of the previous one, and the number of quantizers controlling the overall bitrate. This induces a hierarchical token structure, where tokens from finer RVQ levels contribute less to the perceptual quality, allowing for efficient factorizations and approximations of the joint distribution of the token sequence. Hence, the models and decoding schemes should take this special structure of the input into account for efficient training and inference.

In this work, we present \ours{}, a method for efficient and high-quality audio generation. \ours addresses the problem of generating long audio token sequences by relying on: i) an architecture adapted to the hierarchical structure of the audio tokens, ii) a parallel, non-autoregressive, confidence-based decoding scheme inspired by MaskGIT~\citep{chang2022maskgit} for residual vector-quantized token sequences. 

\ours{} relies on a bidirectional attention-based Conformer~\cite{conformer} that is trained to predict masked audio tokens produced by \soundstream given a conditioning signal such as the semantic tokens of \audiolm\citep{audiolm}. On the input side, it sums up the embeddings of the tokens corresponding to the same \soundstream frame, such that the internal sequence length for the self-attention is identical to the number of \soundstream frames, and independent of the number of quantizers in the RVQ. The output embeddings are then processed by separate heads per RVQ level to predict the masked target tokens.  
At inference time, given the conditioning signal, \ours{} starts with all audio tokens masked out, and fills in the masked tokens RVQ level-by-level over several iterations, predicting multiple tokens in parallel during a single iteration within a level. To support this inference scheme, we propose a masking scheme for training that mimics the inference procedure. 

We demonstrate that \ours can serve as \audiolm's acoustic generator, replacing both \audiolm's stage two (coarse acoustic model) and stage three (fine acoustic model). \ours produces audio two orders of magnitude faster than \audiolm's hierarchical autoregressive acoustic generator with matching quality and improved consistency in terms of speaker identity and acoustic conditions. Furthermore, we show that \ours, coupled with the text-to-semantic modeling stage of \spear\citep{speartts}, can synthesize high-quality, natural dialogues, allowing one to control the spoken content (via transcripts), speaker voices (via short voice prompts) and speaker turns (via transcript annotations). 
When synthesizing dialogues of 30 seconds, we measure a runtime of 2 seconds on a single TPU-v4~\citep{tpuv4}.

\begin{figure*}[t]
    \centering
    \includegraphics[width=0.7\textwidth]{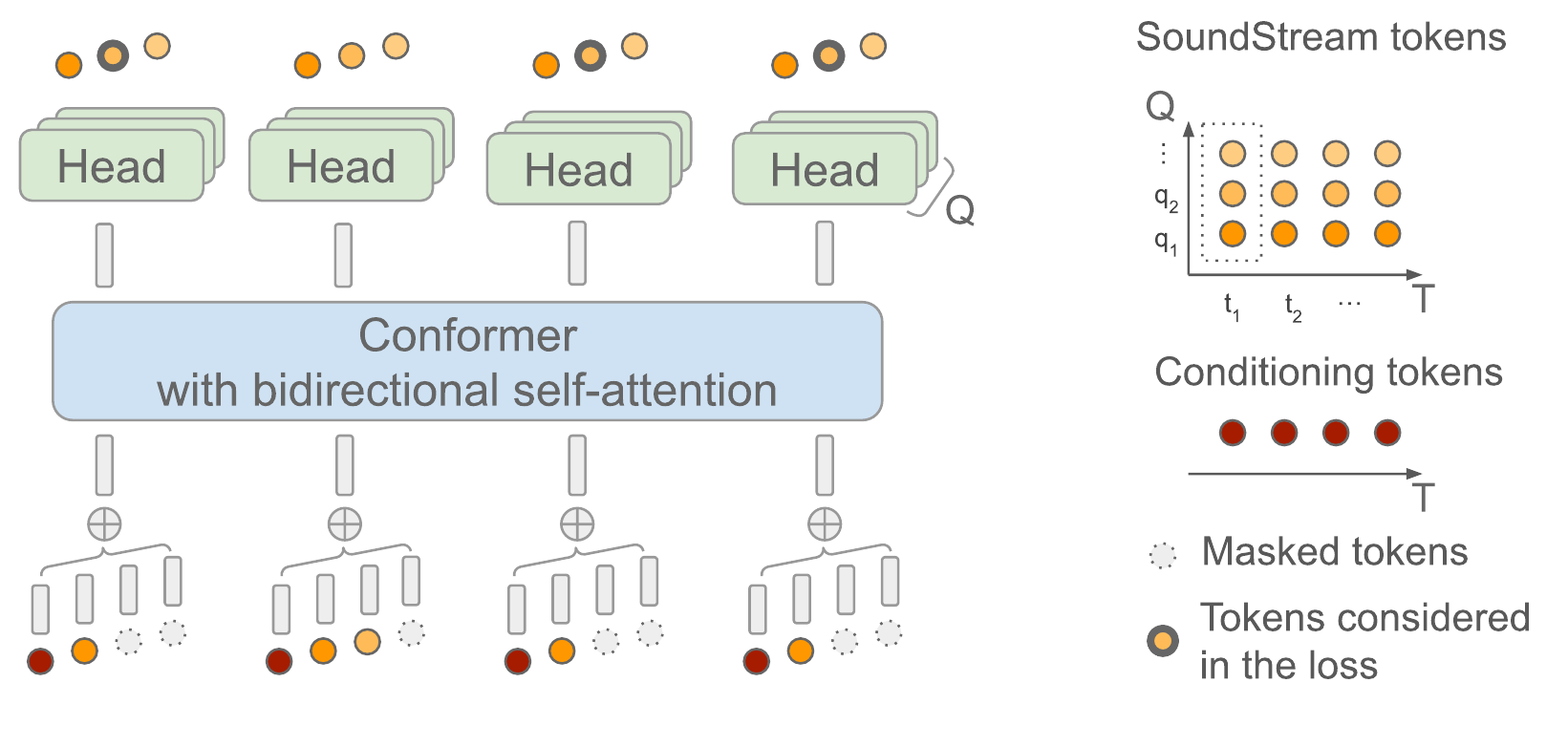}
    \vspace{-2mm}
    \caption{\ours architecture and masking scheme for training (without prompting). The model reduces the input sequence length by summing up the embeddings of the tokens corresponding to the same \soundstream frame. During training, an RVQ level $q$ is sampled ($q = 2$ out of $Q=3$ levels in the figure), and a subset of randomly sampled tokens at level $q$ are masked together with all tokens at RVQ levels $q+1,\dots,Q$. The loss is computed only on the masked tokens at level $q$.  \label{fig:model-architecture}}
    \vspace{-2mm}
\end{figure*}

\section{Related work}
\paragraph{Modeling the tokens of neural audio codecs.} Unsupervised speech embeddings \citep{wav2vec2, hubert, w2vbert} have provided a low-framerate representation of the underlying signal which remains rich enough after discretization for language models to generate intelligible speech from a specific speaker as a sequence of tokens~\citep{lakhotia2021generative}. Neural audio codecs \citep{soundstream,defossez2022high}, with their ability of reconstructing high-quality audio at very low bitrates, subsequently allowed for extending discrete modeling to audio signals as diverse as multi-speaker speech and piano \citep{audiolm,speartts}, music \citep{musiclm} or sound effects \citep{audiogen}. In particular, \audiolm{} \citep{audiolm} introduces a hierarchical sequence-to-sequence approach where high-level semantic tokens  are generated as an intermediate representation, which is then used as a conditioning signal for predicting tokens of a SoundStream \citep{soundstream} codec. While this hierarchical approach has demonstrated remarkable results for speech \citep{speartts} and music modeling \citep{musiclm, singsong}, the computational cost of modeling flattened SoundStream tokens with self-attention scales quadratically with the sequence length and thus the bitrate of the neural codec, preventing these models from generating long-form audio with high quality. \ours{} alleviates this issue by modeling the multi-level tokens of the neural codec in parallel, inducing a two-order of magnitude speed-up over autoregressive modeling and unlocking the ability to scale audio generation abilities both in quality and in sequence length.

\paragraph{RVQ-aware architectures.} A common design choice for modeling RVQ token sequences is to sum the embeddings corresponding to the same RVQ input embedding (frame) in order to reduce the sequence length.
Operating on such sequences, AudioGen~\citep{audiogen} proposes a Transformer with $Q$ separate heads for the different RVQ levels, predicting the tokens for an RVQ frame in parallel. While providing a significant speedup for inference, the authors found that, for text-to-audio generation, this approach has an inferior performance compared to modeling the token sequence of a neural audio codec with similar bitrate and reconstruction quality, but with a single level of quantization.

\looseness=-1
VALL-E~\citep{valle} instead relies on a hybrid approach, where the tokens corresponding to the first RVQ level are predicted autoregressively, and the subsequent levels are produced non-autoregressively. The latter is achieved by a model that sums up the embeddings from the same RVQ input frame, and applies bidirectional self-attention to predict all tokens from RVQ level $q+1$ given all tokens from levels  $1,\dots,q$, the acoustic prompt and the phoneme sequence. During inference, tokens starting from the second level of the RVQ are produced iteratively, performing greedy decoding (choosing the most likely tokens) level-by-level. Level-wise greedy decoding represents the baseline for our method.

Modeling sequences produced by RVQ has been also investigated in domains other than audio. For example, the RQ-Transformer~\citep{lee2022}
also adds up the embeddings corresponding to the same RVQ input frame, but factorizes the full joint distribution efficiently with a spatial and a depth Transformer, for modeling autoregressively the RVQ frames and tokens within the frames, respectively. While it has not been demonstrated yet, this approach, potentially coupled with parallel decoding schemes, is a promising future avenue for audio generation.

\textbf{Parallel decoding.} In order to improve the inference time and to allow bidirectional non-causal attention on the input sequence, parallel decoding schemes have been proposed for text~\citep{nar-nmt,maskpredict}, image~\cite{chang2022maskgit} and video generation~\citep{villegas2022phenaki}. Of particular relevance to our work is the parallel, iterative sampling scheme of MaskGIT~\cite{chang2022maskgit}. During inference time, MaskGIT starts from masked tokens, and in each round, predicts a portion of the tokens based on confidence scores. The portion of the predicted tokens in each round is controlled by a schedule, and usually progressively increases over the iterations --- once predicted, the tokens are treated as fixed. Our proposed decoding scheme can be seen as the extension of MaskGIT's decoding to token sequences produced by residual quantization.

\section{Method}

\looseness=-1
\ours receives as input a sequence of discrete tokens representing the conditioning signal and produces as output a sequence of \soundstream tokens, which can be decoded back to audio waveforms. We assume that the conditioning signal is time-aligned with the \soundstream frames or can be upsampled to the same rate. Such a conditioning signal is, for example, the semantic token sequence used in \audiolm, \spear or MusicLM, which makes our method a drop-in replacement for the acoustic generators of these models. We leave the extension to other types of conditioning signals via cross-attention or to unconditional sampling for future work, and focus our presentation of \ours as the acoustic generator within \audiolm, replacing both \audiolm's coarse and fine acoustic modeling stages.

\subsection{Architecture}
\looseness=-1
The architecture of the model is illustrated in Figure~\ref{fig:model-architecture}. At the input side, we interleave the time-aligned conditioning tokens with the \soundstream tokens at the frame level, embed the resulting sequence, sum the embeddings corresponding to the same frame, including the embedding of the conditioning token, and pass the resulting continuous embeddings to a Conformer. Consequently, the sequence length for bidirectional self-attention in the Conformer is determined by the number of \soundstream frames (typically 50 per second), and thus is independent of the number of RVQ levels $Q$, allowing one to handle audio with length on the order of minutes. At the output side, we use $Q$ dense layers as heads to produce the target \soundstream tokens.
 
\subsection{Masking}
\looseness=-1
For designing our masking and decoding, we extend the masking and confidence-based parallel decoding scheme of MaskGIT~\citep{chang2022maskgit} to token sequences produced by RVQ. At a high level, our approach can be seen as following the strategy of \citet{chang2022maskgit} per RVQ level in a coarse-to-fine order. The coarse-to-fine ordering is of particular importance, since it not only respects the conditional dependencies between levels of the RVQ hierarchy, but also exploits the conditional independence of tokens from finer levels given all tokens from coarser levels. The tokens of finer levels are responsible for local, fine acoustic details and can thus be sampled in parallel without a loss of audio quality.

\looseness=-1
We design our masking scheme for training accordingly. To enable voice prompting, we randomly sample a timestep $t\in \{1,\dots,T\}$, where $T$ denotes the maximum sequence length,  and we do not mask any tokens before this timestep. The conditioning tokens are never masked. Let $Y\in\{1,\dots, C\}^{T\times Q}$ denote the \soundstream tokens, where $C$ indicates the codebook size used in each RVQ level out of the $Q$ levels. Our masking scheme proceeds as follows:
\begin{itemize}[itemsep=1pt,topsep=0pt]
\item Sample the prompt delimiter timestep $t\sim \mathcal{U}\{0, T-1\}$;
\item Sample the current RVQ level $q\sim \mathcal{U}\{1, Q\}$;
\item Sample the mask $M\in\{0, 1\}^T$ according to a cosine schedule~\citep{chang2022maskgit} for level $q$, i.e., sample the masking ratio $p=\cos(u)$ where $u\sim \mathcal{U}[0, \pi/2]$, and sample iid $M_i \sim \textup{Bernoulli}(p)$.
\item Mask the selected non-prompt tokens at the current RVQ level $q$ (mask $Y_{t',q}$ if $M_{t'}=1$ and $t'>t$) and all non-prompt tokens at finer RVQ levels ($Y_{>t, >q}$).
\end{itemize}
Given a masked token sequence, we train the model with cross-entropy loss with the ground-truth tokens as target, where the loss is only calculated on the masked tokens within the $q$-th RVQ level. An example of this masking scheme is illustrated in Figure~\ref{fig:model-architecture}, with $T=4$, $Q=3$, $t=0$ and $q=2$.

\subsection{Iterative Parallel Decoding}
\looseness=-1
Given a conditioning signal, our decoding scheme starts with all \soundstream tokens masked out except for the ones of the prompt (if provided). Then, it proceeds to sampling the tokens RVQ level-wise in a coarse-to-fine order, only proceeding to level $q+1$ when all tokens for levels $1,\dots,q$ have been sampled. Within an RVQ level, we use the confidence-based sampling scheme of~\citet{chang2022maskgit}. Namely, we perform multiple forward passes, and at each iteration $i$, we sample candidates for the masked positions, retaining $p_i$ of them based on confidence scores, where $p_i$ follows a cosine schedule. Compared to~\citet{chang2022maskgit}, we use greedy decoding instead of confidence-based sampling for the last iteration within each RVQ level, which we found to improve the perceived audio quality.

\looseness=-1
Performing the decoding RVQ level-wise makes it possible to exploit the conditional independence assumption in finer levels, namely that multiple finer tokens can be sampled in parallel since they represent local, fine acoustic details. This implies that we can decrease the number of forward passes significantly as we progress to finer RVQ levels during decoding.

\section{Experiments}

\subsection{Model Training and Inference Setup}

In our experiments, we rely on a \soundstream codec that produces 50 frames per second and uses an RVQ with $Q=12$ levels, with 1024 codebook size per level, resulting in a bitrate of $50 \cdot 12 \cdot \log_2 1024 = 6000$ bps. We use the semantic tokens of AudioLM as conditioning, which originate from w2v-BERT~\citep{w2vbert} embeddings quantized with $k$-means with 1024 cluster centers. These tokens have a rate of 25 tokens per second, so we duplicate them to match the framerate of \soundstream.

\looseness=-1
We use a Conformer with 350M parameters, with 12 layers, 16 attention heads, embedding size and model dimension of 1024, feedforward dimension of 4096, convolution kernel size of 5, and rotary positional embeddings~\citep{su2021roformer}. During decoding, we use $(16, 1, 1,\dots, 1)$ iterations for the RVQ levels, that is, 16 iterations in the first level and greedily choosing the tokens with the highest probability in the subsequent levels, level-by-level. This strategy results in 27 forward passes with the model to predict 30 seconds of audio, or the equivalent of $30 \cdot 50 \cdot 12 = 18000$ \soundstream tokens.
We train the model on LibriLight~\citep{librilight} (60k hours), with 10 epochs over the data, sampling random windows of length between 0 and 30 seconds from each example. 

\begin{table*}[t]
\caption{Comparing intelligibility, quality, voice preservation, and acoustic consistency of \audiolm's acoustic generator and \ours. We report metric values for the `short' (4-10 s), `mid' (10-20 s), and `long' (20-30 s) splits of LibriSpeech test-clean separately.
\ours matches \audiolm's acoustic generator in terms of audio quality, and outperforms it in terms of speech intelligibility and acoustic consistency. 
\label{table:results}}
\resizebox{\textwidth}{!}{
\centering
\begin{tabular}{ccccccccccccccccccc}
\toprule
& 
\multicolumn{3}{c}{\textbf{WER}$\downarrow$} & \multicolumn{3}{c}{\textbf{CER}$\downarrow$} &
\multicolumn{3}{c}{\begin{tabular}[c]{@{}c@{}}\textbf{Audio}\\ \textbf{quality}\end{tabular}$\uparrow$} & 
\multicolumn{3}{c}{\begin{tabular}[c]{@{}c@{}}\textbf{Voice}\\ \textbf{preservation}\end{tabular}$\uparrow$} &
\multicolumn{3}{c}{\begin{tabular}[c]{@{}c@{}}\textbf{Acoustic}\\ \textbf{consistency}\end{tabular}$\uparrow$} \\

& \tiny short & \tiny mid & \tiny long & \tiny short & \tiny mid & \tiny long & \tiny short & \tiny mid & \tiny long & \tiny short & \tiny mid & \tiny long & \tiny short & \tiny mid & \tiny long \\
\midrule
\begin{tabular}[c]{@{}c@{}}Original\\ \soundstream rec.\end{tabular}  & $2.62$ & $1.95$ & $2.20$   &  $0.89$ & $0.55$ & $0.69$  &  $3.72$ & $3.91$ & $3.99$ & $0.63$ & $0.65$ & $0.66$ & $0.97$ & $0.95$ & $0.93$\\
\midrule
& \multicolumn{15}{c}{\textit{Without a speaker prompt}} \\
AudioLM\vspace{1mm} &    $4.65$ & $3.59$ & $4.79$      &    $2.15$ & $1.57$ & $2.30$   & $3.93$ &  $4.04$ & $4.08$ &    $-$ & $-$  & $-$    & $-$    & $-$    & $-$     \\
\ours                             &   \textbf{3.48} & \textbf{2.55} & \textbf{3.33} &  \textbf{1.39} & \textbf{0.89} & \textbf{1.29 } & \textbf{4.01} & \textbf{4.16} & \textbf{4.20} & $-$ & $-$    & $-$       & $-$    & $-$  & $-$   \\
\midrule
& \multicolumn{15}{c}{\textit{With a speaker prompt}} \\
AudioLM &  $3.77$ & $3.40$ & $3.75$   & $1.50$ & $1.47$ & $1.54$ & \textbf{3.91} & \textbf{4.06} & $4.10$  &  $0.46$ & $0.48$ & $0.48$ & \textbf{0.96} & $0.91$ & $0.86$ \\
\ours  & \textbf{2.99} & \textbf{2.43} & \textbf{3.36} & \textbf{1.10} & \textbf{0.81} & \textbf{1.24} & 
3.81 & \textbf{4.05} & \textbf{4.15} & \textbf{0.57} & \textbf{0.59} & \textbf{0.59} & \textbf{0.96} & \textbf{0.94} & \textbf{0.91} \\ \bottomrule
\end{tabular}
}
\end{table*}

\subsection{Speech Intelligibility, Audio Quality, Voice Preservation and Acoustic Consistency}

\looseness=-1
In a series of subjective evaluation experiments, \citet{audiolm} and \citet{speartts} have shown that the acoustic generation stage of AudioLM produces audio with quality indistinguishable from the quality of the ground-truth samples. Hence, we consider AudioLM's hierarchical acoustic generation stages (coarse and fine stages) as a baseline in our experiment. One important difference compared to the AudioLM acoustic stage experiments of \citet{audiolm} is that we require the conditioning signal to be time-aligned with the \soundstream frames, whereas the experiments of \citet{audiolm} use semantic tokens with duplicates removed. For a fair comparison, so that both methods use the same conditioning, we repeat the acoustic stage experiments of \citet{audiolm} without removing duplicate semantic tokens, thus making the conditioning stronger. We also use the same \soundstream codec for both methods.

\paragraph{Speech intelligibility.} We quantify speech intelligibility by measuring the word error rate (WER) and character error rate
(CER) of the generated audio after transcribing it with ASR. The generation is conditioned on the ground-truth semantic tokens from LibriSpeech test-clean split~\citep{librispeech}. We perform these experiments both in the unprompted setup, where the methods can randomly sample speakers, and in the prompted setup, where the methods should respect the speaker identity provided in the form of ground-truth \soundstream tokens corresponding to the first 3-seconds. We use a Conformer Transducer-L~\citep{conformer} ASR model for transcription. 

We report the results separately on short (4-10 seconds), medium (10-20 seconds) and long (20-30 seconds) segments.  Since \audiolm's acoustic generation stage is autoregressive in the flattened token sequence of the first 4 RVQ levels, it is prohibitively expensive to generate 30 seconds in a single pass. Hence, we generate segments longer than 10 seconds with a slide-and-prompt approach, where we generate 10-second chunks and use the last 3 seconds as the prompt for the next chunk. The results in Table~\ref{table:results} show that \ours improves over \audiolm's acoustic generation significantly in terms of WER and CER on all splits, both for the prompted and unprompted scenario.

\paragraph{Voice preservation.} Next, we measure the capability of \ours{} to maintain the speaker identity of the prompt. To do so, we extract speaker embeddings from the prompt and the generated audio and compute their cosine similarity. As in~\citet{valle} and~\citet{speartts}, we compute embeddings with a publicly available\footnote{\url{https://github.com/microsoft/UniSpeech/tree/main/downstreams/speaker\_verification\#pre-trained-models}} speaker verification system based on WavLM~\citep{wavlm}.
Table 1 shows that \ours{} significantly outperforms the \audiolm baseline. 

\begin{figure}[t]
    \centering
    \includegraphics[width=0.87\columnwidth]{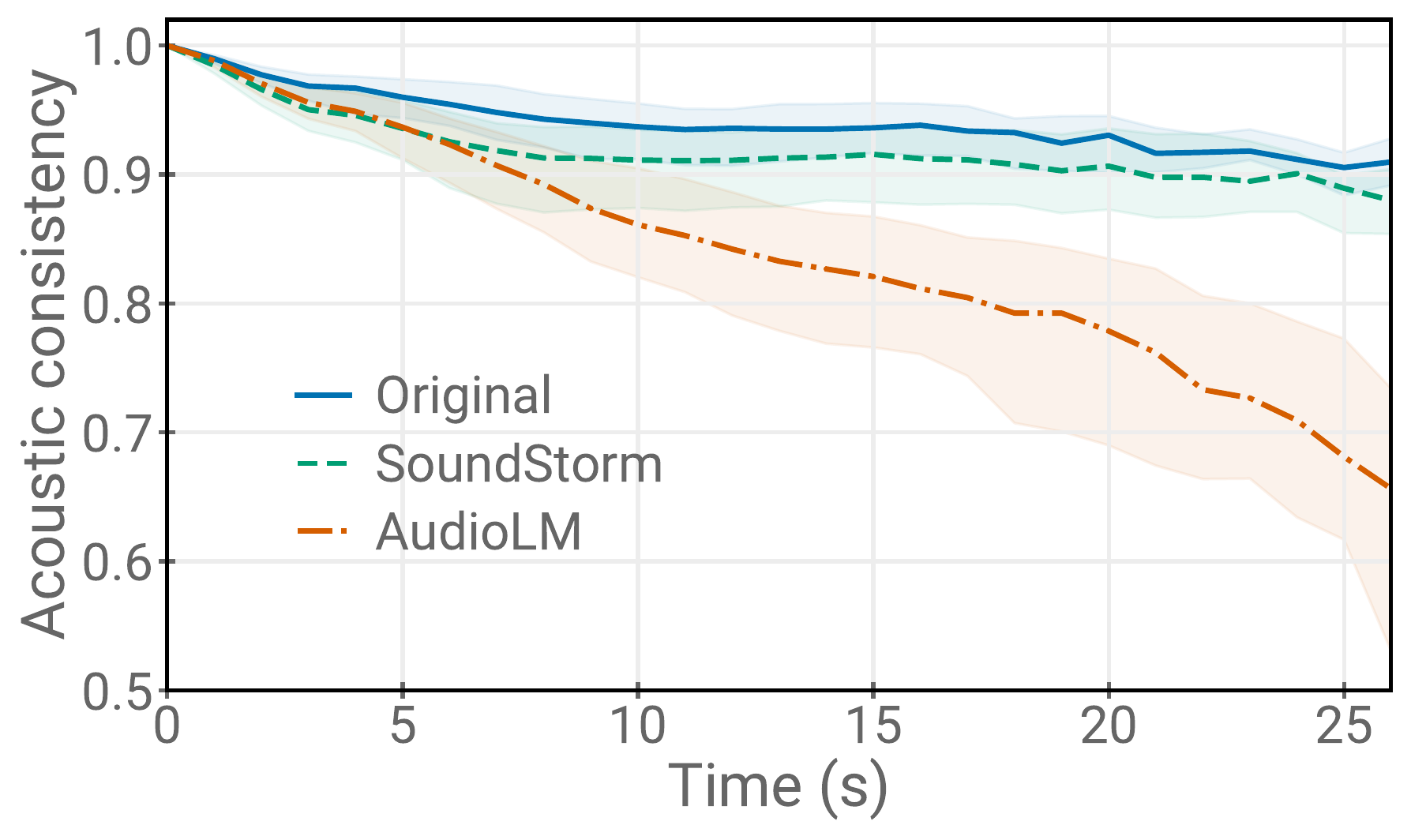}
    \vspace{-3mm}
    \caption{Acoustic consistency between the prompt and the generated audio for the samples in the `long' split of LibriSpeech test-clean. The shaded area represents the inter-quartile range.  \label{fig:acoustic-consistency}}
    \vspace{-3mm}
\end{figure}

\paragraph{Acoustic consistency drift.} In the case of long generation, it is interesting to measure to what extent the acoustic properties of the prompt (e.g., speaker identity, recording conditions) are preserved along time in the generated audio. To this end, we train a model capable of assessing whether two short segments come from the same recording. Specifically, we consider a speech corpus in which each example contains a recording from a single speaker (e.g., LibriLight) and extract two random, non-overlapping crops with duration of 2-5 seconds. Then, from each crop, we extract a sequence of  embeddings from an intermediate layer of a BEST-RQ model~\cite{bestrq} pre-trained on the same corpus. We use layer 4 in our experiments. Each sequence of embeddings is fed to a model that consists of a Conformer layer, followed by global average pooling along the temporal dimension and a linear projection layer, so that each crop is represented with a single 1024-dimensional embedding. We train this model using the same contrastive loss as~\citet{clip}. 

\looseness=-1
To measure acoustic consistency drift, we compute the cosine similarity between the embedding computed from the prompt of 3 seconds, and the embeddings computed from subsequent crops of 3 seconds along the temporal axis. As an illustration, Figure~\ref{fig:acoustic-consistency} shows the measured drift on the `long' split of LibriSpeech test-clean. We observe that for the audio generated by \ours{}, the acoustic consistency score is close to the one measured for the original samples, while for \audiolm we observe a more significant drift over time. Table~\ref{table:results} reports the average acoustic consistency scores on the `short', `mid' and `long' splits, where the averaging is done over non-overlapping crops. Unsurprisingly, the improvement is more evident for longer audio samples.

\paragraph{Audio Quality.} We use a MOS estimator similar to DNSMOS~\citep{dnsmos} to estimate the perceived audio quality of the generated samples. The results in Table~\ref{table:results} show that, according to the MOS estimator, \ours is on par with \audiolm's acoustic generator, which in turn has been shown to match the quality of the ground-truth audio in the subjective studies of \citet{audiolm} and \citet{speartts}.

\begin{figure}[t]
    \centering
    \includegraphics[width=0.9\columnwidth]{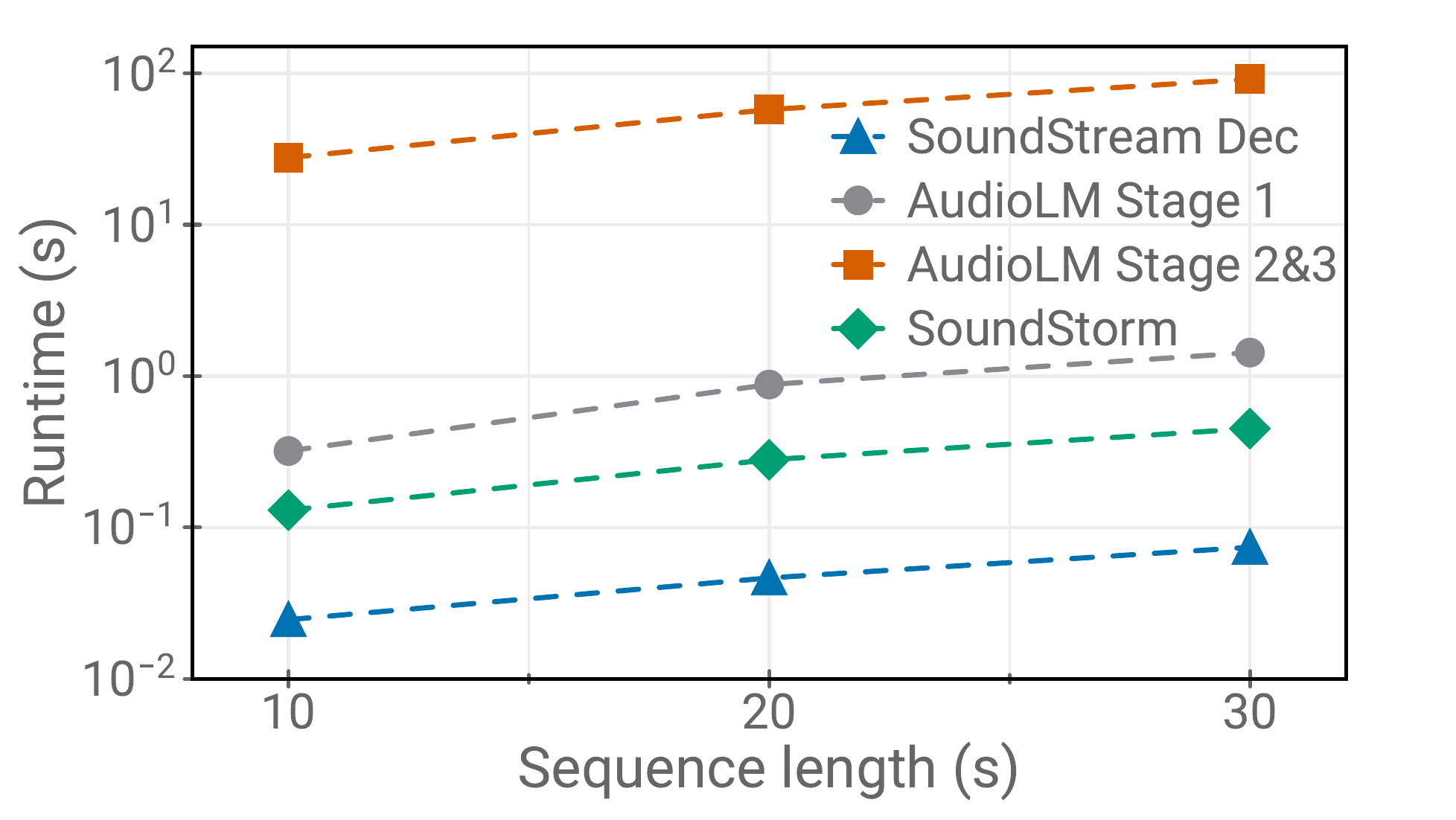}
    \vspace{-2mm}
    \caption{Runtimes of \soundstream decoding, \ours and different stages of  \audiolm on a TPU-v4. \label{fig:runtimes}}
\end{figure}

\begin{figure}[t]
    \centering
    \includegraphics[width=0.855\columnwidth]{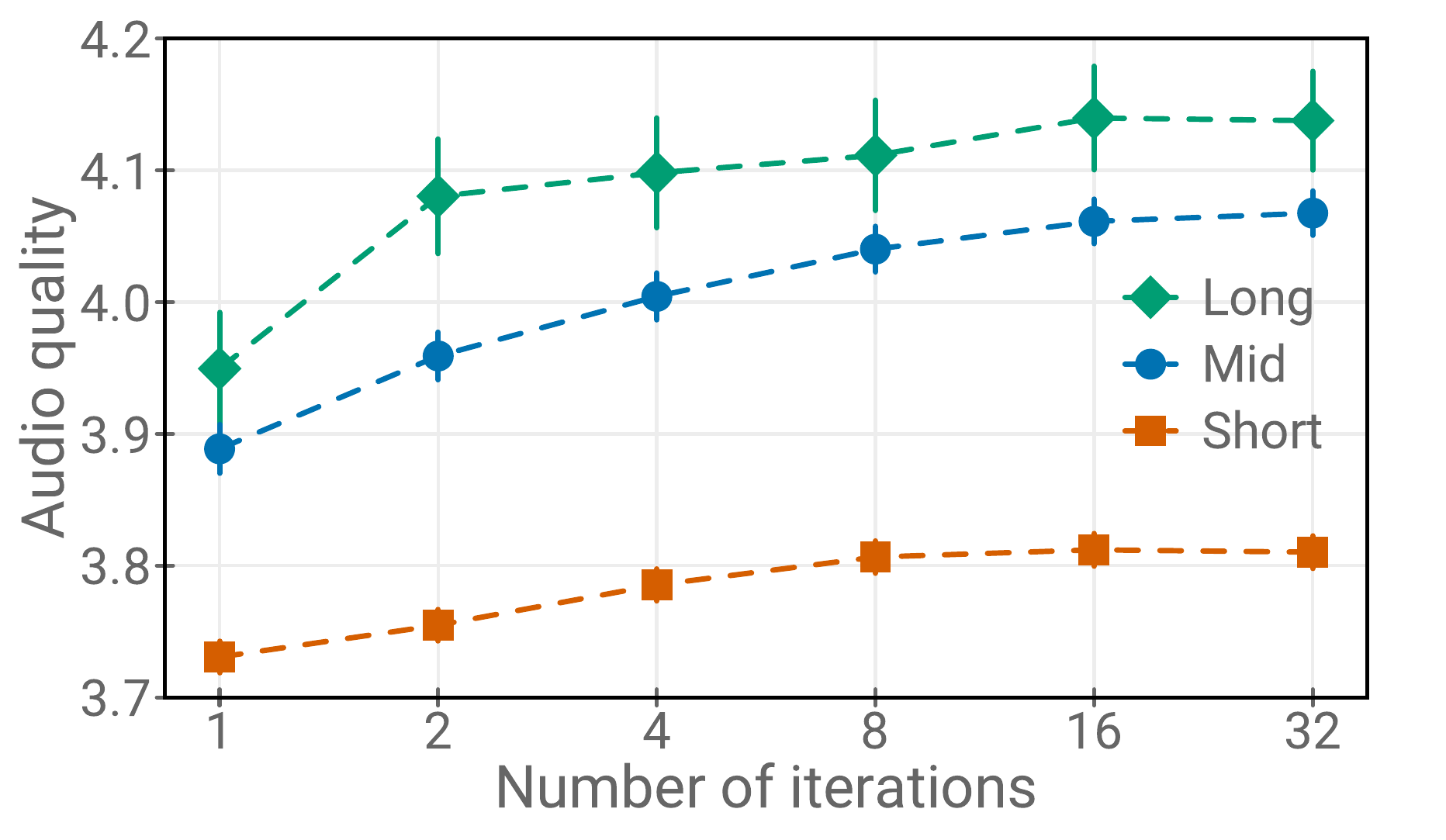}
    \vspace{-2mm}
    \caption{Audio quality with respect to the number of decoding iterations in the first RVQ level. \label{fig:iter-ablation}}
    \vspace{-2mm}
\end{figure}

\subsection{Runtime and Ablations}

\paragraph{Runtime.} We measure the runtime of the different methods to generate up to 30 seconds of audio on a single TPU-v4. Figure \ref{fig:runtimes} shows that \ours can generate audio two orders of magnitude faster than AudioLM's acoustic generator (``\audiolm Stage 2\&3'' in the figure), with a real time factor of 0.017, including decoding to waveform by the \soundstream decoder. We also measure the runtime of the semantic generation stage of \audiolm (``\audiolm Stage 1'' in the figure), and conclude that by coupling the semantic generation stage with \soundstream, we can generate 30 seconds of speech continuation within 2 seconds (1.4 seconds for semantic generation, 0.5 seconds for \ours and 0.1 seconds for \soundstream decoding).

\paragraph{Number of decoding steps.} In the former experiments, we used 16 iterations for decoding the first RVQ level, and performed greedy decoding for the subsequent levels. We now investigate the effect of the number of decoding iterations for the different RVQ levels.

To achieve this, we repeat the speaker-prompted experiment on LibriSpeech test-clean with different number of decoding iterations in the first level. Figure \ref{fig:iter-ablation} shows that, according to the audio quality estimator, our strategy of using 16 iterations achieves an increase of 0.1-0.2 in the quality score compared to level-wise greedy decoding, whereas further increasing the number of iterations does not improve the score. The artifacts produced by the greedy strategy are clearly perceivable; we provide samples produced by this strategy on the accompanying webpage.

\looseness=-1
We have also experimented with increasing the number of iterations for RVQ levels 2-12, and we have found no statistically significant improvement in the audio quality score when synthesizing samples from LibriSpeech test-clean. We note that this observation is in line with the decoding strategy of \citet{valle}, which is autoregressive in the first RVQ level, and level-wise greedy beyond. We hypothesize that performing multiple iterations on finer levels becomes relevant when generating audio beyond speech, where the semantic tokens and first RVQ level \soundstream tokens will not capture all the important acoustic details.

\section{Dialogue Synthesis}
\looseness=-1
Spoken dialogue synthesis is an important application where maintaining speaker identities over multiple speaker turns and long time spans is of paramount importance. This makes it an ideal use-case for \ours. In this section, we show that, by coupling \ours with a text-to-semantic token model, we can synthesize high-quality multi-turn dialogues with duration of up to 30 seconds. This approach is similar to the one of \spear~\citep{speartts}.

We collected a corpus of approximately 100,000 hours of dialogues, segmented into 30-second chunks. By running an off-the-shelf ASR system, we generated a transcript for each segment. In these transcripts, speaker turns are annotated with a turn marker symbol in the appropriate location in the transcript. To extract semantic tokens, we train a 0.6B parameter BEST-RQ~\citep{bestrq} on this dataset and fit $k$-means with 4096 cluster centers to the activations of layer 13 of the BEST-RQ model. This results in 25 semantic tokens per seconds, with a codebook size of 4096. We train a \soundstream codec operating at 24 kHz, producing 50 frames per second with 12 RVQ on this corpus.

To model the text-to-semantic token mapping, we train a ByT5-large Transformer~\cite{byt5}. This is an encoder-decoder model with 36 encoder and 12 decoder layers, embedding size of 1536 and feed-forward dimension of 3840. In total, the model has 1.2B parameters. We only train the decoder and use a text-pretrained encoder from a published ByT5 checkpoint~\cite{byt5}. 
As input, this model takes a byte-level representation of the text and predicts non-deduplicated semantic tokens. Decoding is done by temperature sampling, with temperature of 0.9 and top-k set to 125. Due to the modest size of the decoder and the short target sequence length (only 750 semantic tokens for 30 seconds), running inference with this model takes 1.4 seconds on a TPU-v4. 

We train both the text-to-semantic model and \ours on the dialogue corpus for 10 epochs.
For inference, we recorded short exchanges from speakers not seen during training, serving as prompts for our models, and we created text transcripts that are continuations of the prompts.
We then feed the transcripts to the text-to-semantic model, the output of which is fed to \ours, while using the speaker prompts for both stages. 

\looseness=-1
We find that this approach generates high-quality, natural dialogue sequences, generating disfluencies at the occurrence of filler words in the transcript, and allowing for controlled speaker turns through the insertion of the turn marker symbols in the transcript. The total runtime for synthesizing a segment of 30 seconds segment is 2 seconds. We invite the reader to listen to the generated samples on the accompanying webpage.\footnote{\url{https://google-research.github.io/seanet/soundstorm/examples/}}

\section{Conclusion}
\looseness=-1
In this paper we present \ours{}, a model that can synthesize high-quality audio from discrete conditioning tokens efficiently. When compared to the acoustic generator of \audiolm{}, \ours{} is two orders of magnitude faster and achieves higher temporal consistency when generating long audio samples. By combining a text-to-semantic token model similar to \spear{} with \ours{}, we can scale text-to-speech synthesis to longer contexts and generate natural dialogues with multiple speaker turns, controlling both the voices of the speakers and the generated content.

\section{Broader Impact}
\looseness=-1
\ours is a model for high-quality, efficient generation of neural audio codec-derived representations of audio. In this work, we use it as a replacement for the acoustic generation pipeline of \audiolm and \spear. We acknowledge that the audio samples produced by the model may be influenced by the biases present in the training data, for instance in terms of represented accents and voice characteristics. In our generated samples, we demonstrate that we can reliably control speaker characteristics via prompting. However, a more thorough analysis of any training data and its limitations would be an area of future work in line with our responsible AI principles.

In turn, the ability to mimic a voice can have numerous malicious applications, including bypassing biometric identification and for the purpose of impersonation. Thus, it is crucial to put in place safeguards against potential misuse: to this end, we have verified that the audio generated by \ours remains detectable by a dedicated classifier (98.5\% using the same classifier as~\citet{audiolm}). Hence, as a component of a larger system, we believe that \ours would be unlikely to introduce additional risks to those discussed previously by~\citet{audiolm} and~\citet{speartts}. At the same time, relaxing the memory and computational requirements of AudioLM would make research in the domain of audio generation more accessible to a wider community. In the future, we plan to explore other approaches for detecting synthesized speech, e.g., audio watermarking, so that any potential product usage of this technology strictly follows our responsible AI principles.

\section{Acknowledgements}
\looseness=-1
The authors would like to thank Aren Jansen and RJ Skerry-Ryan for technical discussions and helpful feedback; Jelena Anti\'c, Brian McWilliams, Paul Rubenstein, Michael Dooley and other voice actor volunteers for donating speech samples. 

\bibliography{bibliography}
\bibliographystyle{icml2023}


\end{document}